\documentclass[11pt]{article}
\usepackage{moriond,epsfig}

\newcommand{\DMR}{DMR}

\newcommand{\Boomerang}{BOOMERANG}
\newcommand{\Maxima}{MAXIMA}
\newcommand{\DASI}{DASI}

\newcommand{\cbi}{CBI}

\def\PRD{{Phys. Rev.} D}

\def\apj{{Ap.J.}}

\def\be{\begin{equation}}
\def\ee{\end{equation}}
\def\bea{\begin{eqnarray}}
\def\eea{\end{eqnarray}}

\begin{document}
\vspace*{4cm}
\title{Cosmological Parameters from CMB measurements with the CBI}

\author{ C.~R.~Contaldi${}^1$, J.~R.~Bond${}^1$, D.~Pogosyan${}^2$, B.~S.~Mason${}^{3,4}$,
S.~T.~Myers${}^5$, T.~J.~Pearson${}^3$, U.~L.~Pen${}^1$, S.~Prunet${}^{1,6}$, A.~C.~Readhead${}^3$,
M.~I.~Ruetalo${}^{1,7}$, J.~L.~Sievers${}^3$, J.~W.~Wadsley${}^8$, P.~J.~Zhang${}^{1,7}$}

\address{${}^1$Canadian Institute for Theoretical Astrophysics\\ 60 St. George
Street,  Toronto Ontario M5S 3H8\\ 
${}^2$ Physics Department, University of Alberta, Edmonton, Canada\\
${}^3$ California Institute of Technology, 1200 East California Boulevard, Pasadena, CA 91125\\
${}^4$ National Radio Astronomy Observatory, P.O. Box 2, Green Bank, WV 24944\\
${}^5$ National Radio Astronomy Observatory, P.O. Box O, Socorro, NM 87801\\
${}^6$ Institut d'Astrophysique de Paris, 98bis Boulevard Arago, F 75014 Paris, France \\
${}^7$ Department of Astronomy and Astrophysics, University of
Toronto, 60 St. George Street, Toronto Ontario M5S 3H8 \\
${}^8$ Department of Physics and Astronomy, McMaster University,
Hamilton, ON L8S 4M1, Canada
}

\maketitle\abstracts{ We derive cosmological parameters from the CBI
measurements of the Cosmic Microwave Background (CMB) angular power
spectrum. Our results provide an independent confirmation of the
standard $\Omega_{\rm tot}=1$ $\Lambda$CDM model within the adiabatic,
inflationary paradigm. Above $\ell=2000$ the observations show
evidence of power in excess of that expected in the standard
models. We use hydrodynamical simulations to show how
Sunyaev-Zeldovich Effect (SZE) may account for the excess power for
models with fluctuation amplitude $\sigma_8\sim 1$ which is in the
high end of the range allowed by the primary CMB observations.}

\section{Introduction}

Increasingly accurate measurements of the angular power spectrum of
the Cosmic Microwave Background (CMB) have begun to constrain
cosmological models of structure formation. Previous experiments such
as \Boomerang\cite{Netterfield01}, \DASI\cite{DASI01} and
\Maxima\cite{Maxima01} have now measured precisely the shape of a
first peak at $\ell \sim 220$, consistent with an $\Omega_{tot}=1$
$\Lambda$CDM universe with adiabatic, inflationary seeded
perturbations. A significant detection of a second peak has also been
established \cite{deBernardis01} with evidence for a third. The \cbi\
observations have now confirmed another important element of the
adiabatic, inflationary paradigm, the damping at high multipoles due
to the viscous drag over the finite width of the last scattering
surface \cite{Sievers02}. B. Mason\cite{Mason02a}, in these
proceedings, gives a description of the \cbi\ instrumental setup and
observing strategy and discusses the power spectra. Here we present the
cosmological parameter fits obtained from the observations and discuss
the nature of the possible excess observed on the smallest angular
scales at $2000<\ell<4000$.

During the year 2000 observing season, the \cbi\ covered three deep
fields of diameter roughly $1^\circ$ \cite{Mason02a,Mason02b}, and
three mosaic regions, each of size roughly 13 square degrees
\cite{Pearson02}. The instrument observes in 10 frequency channels
spanning the band $26-36$GHz and measures 78 baselines
simultaneously. Our power spectrum estimation pipeline is described in
\cite{Myers02} and involves an optimal compression of the ${\cal
O}(10^5)$ visibility measurements of each fields into a coarse grained
lattice of visibility estimators. Known point sources are projected
out of the data sets when estimating the primary anisotropy spectrum by
using a number of constraint matrices. The positions are obtained from
the (1.4GHz) NVSS catalog \cite{Condon98}. When projecting out the
source we use large amplitudes which effectively marginalize over all
affected modes. This insures robustness with respect to errors in the
assumed fluxes of the sources. The residual contribution of sources
below our $S_{1.4}=3.4$ mJy cutoff is treated as a white noise background
with an estimated amplitude of $0.08\pm 0.04$ Jy/sr${^-1}$
\cite{Mason02a,Mason02b}.


\section{Cosmological Parameters}

We fit a set of minimal, adiabatic inflationary models to the CBI data
and a combination of the CBI data and previous data sets. Our model
database spans a seven dimensional grid over the parameters $\omega_b
= \Omega_bh^2, \omega_{\rm cdm}=\Omega_{\rm cdm}h^2, \Omega_{\Lambda},
\Omega_{tot}, n_s, \tau_c$ and a continuous amplitude parameter $\ln
{\cal C}_{10}$. We also maximize over a number of internal parameters
which include the beam and calibration errors for each experiment
\cite{Sievers02}.

We use the offset lognormal approximation $Z_B = \ln(C_B+x_B)$ to the
non-Gaussian distribution of the band powers $C_B$ in fitting the data
to models ${\cal C}_{\ell}$. We obtained marginalized one dimensional
distributions for each parameter $x$ by integrating the measured
likelihood function $L(x,\vec y)$ over the remaining parameters $\vec
y$ multiplied by a set of prior functions $P(x,\vec y)$ as ${\cal L}(x) = \int
P(x,\vec y)L(x,{\vec y})d{\vec y}$. Our quoted $1\sigma$
uncertainties and central values are obtained by
computing the $16\%$, $50\%$ and $84\%$ integrals of the marginalised 1-D
likelihoods in each parameters. We also quote confidence limits for
parameters which can be derived from combinations of the basic parameter
set such the Hubble constant $h$ and the age of the universe. Our
set of priors include weak constraints on $h$ and the age of the
universe, a strong constraint on $h$ \cite{Freedman}, a constraint on
the combination $\sigma_8\Omega_M^{0.56}$ and the shape parameters
$\Gamma\approx\Omega_M h$ derived from Large Scale Structure (LSS)
observations, constraints from SN1a observations \cite{SNa,SNb}
and a flatness prior $\Omega_{tot} = 1$. For detailed discussion of
our priors see \cite{Sievers02}.

Our results for the combination of \cbi\ and \DMR\ data are shown in
Table~\ref{tab:param} for various combinations of priors. Our
parameter fits and best fit models are consistent with previous
results from fits to data at lower multipoles. We have also
carried out a combined analysis of the \cbi\ data with the \Boomerang,
\Maxima\ and \DASI\ data and with a compilation of data predating
April 2001 \cite{Sievers02}. The results are consistent with those
quoted here.

\begin{center}
\begin{table}[th]
\caption{Parameters from CBI+DMR\label{tab:param}}
\vspace{0.4cm}
\begin{tabular}{|l|cccccccc|}
\hline
{Priors}
&{$\Omega_{tot}$}
&{$n_s$}
&{$\Omega_bh^2$}
&{$\Omega_{\rm cdm}h^2$}
&{$\Omega_{\Lambda}$}
&{$h$}
&{Age}
&{$\tau_c$}\\
\hline
{\small wk}
& $0.99^{0.12}_{0.12}$ 
& $1.05^{0.09}_{0.08}$ 
& $0.022^{0.015}_{0.009}$ 
& $0.17^{0.08}_{0.06}$ 
& $0.40^{0.25}_{0.27}$ 
& $0.59^{0.11}_{0.11}$ 
& $13.6^{2.0}_{2.0}$ 
& $0.22^{0.19}_{0.16}$ 
\\
{\small wk+LSS}
& $1.01^{0.09}_{0.06}$ 
& $1.02^{0.11}_{0.07}$ 
& $0.026^{0.014}_{0.010}$ 
& $0.12^{0.03}_{0.03}$ 
& $0.64^{0.11}_{0.14}$ 
& $0.62^{0.12}_{0.12}$ 
& $14.6^{2.0}_{2.0}$ 
& $0.14^{0.22}_{0.11}$ 
\\
{\small wk+SN}
& $1.02^{0.09}_{0.08}$ 
& $1.07^{0.09}_{0.09}$ 
& $0.027^{0.015}_{0.011}$ 
& $0.12^{0.05}_{0.05}$ 
& $0.70^{0.08}_{0.09}$ 
& $0.67^{0.12}_{0.12}$ 
& $14.3^{2.6}_{2.6}$ 
& $0.24^{0.18}_{0.18}$ 
\\
{\small wk+LSS+SN}
& $1.00^{0.10}_{0.06}$ 
& $1.06^{0.09}_{0.08}$ 
& $0.027^{0.014}_{0.011}$ 
& $0.12^{0.04}_{0.04}$ 
& $0.70^{0.07}_{0.07}$ 
& $0.68^{0.12}_{0.12}$ 
& $14.1^{2.3}_{2.3}$ 
& $0.21^{0.20}_{0.15}$ 
\\\hline
{\small Flt+wk}
& (1.00) 
& $1.04^{0.10}_{0.08}$ 
& $0.023^{0.010}_{0.008}$ 
& $0.15^{0.06}_{0.04}$ 
& $0.46^{0.22}_{0.29}$ 
& $0.60^{0.12}_{0.12}$ 
& $13.8^{1.4}_{1.4}$ 
& $0.22^{0.19}_{0.16}$ 
\\
{\small Flt+wk+LSS}
& (1.00) 
& $1.01^{0.10}_{0.07}$ 
& $0.025^{0.010}_{0.008}$ 
& $0.13^{0.02}_{0.01}$ 
& $0.64^{0.10}_{0.13}$ 
& $0.65^{0.12}_{0.12}$ 
& $14.0^{1.2}_{1.2}$ 
& $0.15^{0.17}_{0.11}$ 
\\
{\small Flt+wk+SN}
& (1.00) 
& $1.06^{0.11}_{0.09}$ 
& $0.026^{0.010}_{0.009}$ 
& $0.13^{0.03}_{0.02}$ 
& $0.69^{0.06}_{0.07}$ 
& $0.71^{0.09}_{0.09}$ 
& $13.3^{1.1}_{1.1}$ 
& $0.22^{0.19}_{0.16}$ 
\\
{\small Flt+wk+LSS+SN}
& (1.00) 
& $1.05^{0.09}_{0.07}$ 
& $0.027^{0.009}_{0.009}$ 
& $0.13^{0.02}_{0.01}$ 
& $0.70^{0.05}_{0.06}$ 
& $0.71^{0.08}_{0.08}$ 
& $13.4^{0.9}_{0.9}$ 
& $0.20^{0.16}_{0.14}$ 
\\\hline
{\small Flt+HST}
& (1.00) 
& $1.06^{0.10}_{0.08}$ 
& $0.026^{0.010}_{0.009}$ 
& $0.15^{0.07}_{0.04}$ 
& $0.61^{0.10}_{0.21}$ 
& $0.67^{0.08}_{0.08}$ 
& $13.1^{1.2}_{1.2}$ 
& $0.21^{0.19}_{0.16}$ 
\\
{\small Flt+HST+LSS}
& (1.00) 
& $1.04^{0.08}_{0.07}$ 
& $0.027^{0.009}_{0.008}$ 
& $0.13^{0.02}_{0.01}$ 
& $0.68^{0.05}_{0.07}$ 
& $0.68^{0.07}_{0.07}$ 
& $13.6^{0.8}_{0.8}$ 
& $0.19^{0.15}_{0.13}$ 
\\
{\small Flt+HST+SN}
& (1.00) 
& $1.06^{0.11}_{0.09}$ 
& $0.027^{0.009}_{0.009}$ 
& $0.13^{0.03}_{0.02}$ 
& $0.69^{0.04}_{0.06}$ 
& $0.70^{0.05}_{0.05}$ 
& $13.4^{0.8}_{0.8}$ 
& $0.22^{0.19}_{0.16}$ 
\\
{\small Flt+HST+LSS+SN}
& (1.00) 
& $1.05^{0.08}_{0.07}$ 
& $0.027^{0.009}_{0.009}$ 
& $0.13^{0.02}_{0.01}$ 
& $0.70^{0.04}_{0.05}$ 
& $0.70^{0.05}_{0.05}$ 
& $13.5^{0.6}_{0.6}$ 
& $0.20^{0.15}_{0.14}$ 
\\\hline
\end{tabular}
\end{table}
\end{center}

\section{The CBI Deep Field Excess and the Sunyaev-Zeldovich Effect}

In \cite{Mason02a,Mason02b} we reported the results of deep 
observations of three single fields. The measurements reveal an
apparent excess in the power at multipoles $\ell > 2000$ over standard
adiabatic, inflationary models with a significance of $3.1\sigma$. The
excess is a factor of $4.5$ greater than the estimated contribution
from a background of residual sources and the confidence limit
includes a $50\%$ error in the value for the background flux density.

\begin{figure}
\begin{center}
\psfig{figure=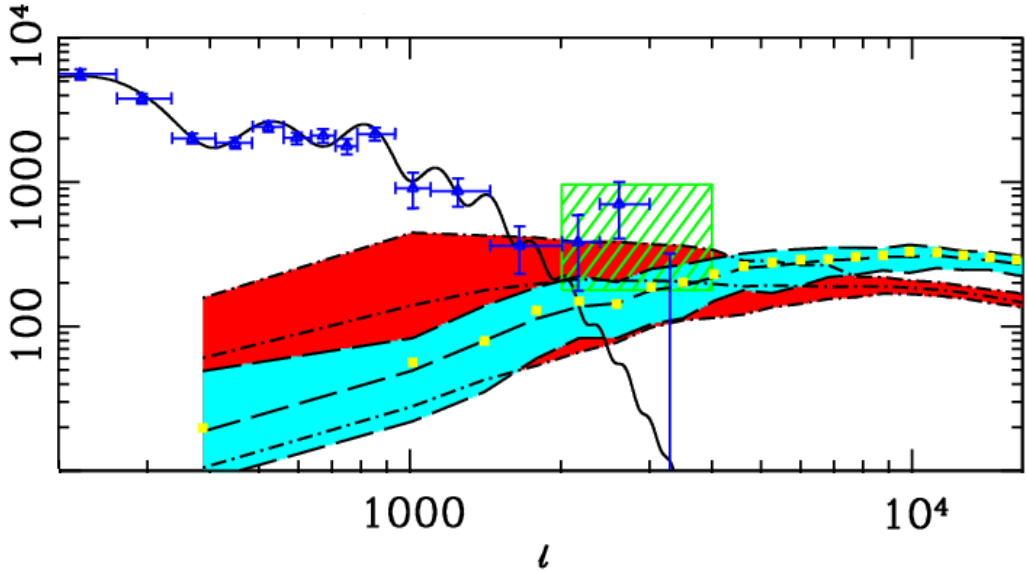,height=3.1in}
\end{center}
\caption{The SZE powers spectrum for $\sigma_8=1$. The triangle (blue) points are an optimal combination of all data. The hatched (green) area represents the 2-sigma confidence region for the high-$\ell$ CBI band. The area bounded by dash-dotted curves (red) cover the results from 20 MMH SZ maps while the area bounded by long-dashed (cyan) line cover the results from the 10 SPH maps. The square (yellow) points are the results of the $\sigma_8=0.9$ 400 Mpc result scaled to $\sigma_8=1.0$ using our empirical relation.
\label{fig:szspectrum}}
\end{figure}

We have considered whether secondary anisotropies from the
Sunyaev-Zeldovich effect may explain the observed excess
\cite{Bond02}. We used four hydrodynamical simulations employing both
Smoothed Particle Hydrodynamics (SPH) and Moving Mesh Hydrodynamics
(MMH) algorithms with rms amplitudes $\sigma_8=1.0$ and $0.9$ to
calculate the expected contribution to the angular power spectrum from
the SZE. We find that both algorithms produce power consistent with
the observed excess for $\sigma_8=1$. In Figure~\ref{fig:szspectrum}
we show the result of the MMH and SPH simulations. The spectra are
obtained by 20 and 10, 4 square degree simulated maps respectively.

The \cbi\ power spectrum estimation pipeline has been tested
extensively using accurate simulations of the observations with the
exact $uv$-coverage and noise characteristics of the observed fields
\cite{Myers02}. We used the SZ maps from the hydrodynamical codes as
foregrounds in our simulations to test the Gaussian assumption
implicit in the bandpower estimation algorithm in the presence of
extended non-Gaussian foregrounds such as the SZE. We find that, for
the amplitudes considered, the pipeline recovers the total power
accurately or the 30 maps considered \cite{Bond02} including at scales
$\ell > 2000$ where the signal is dominated by the SZ foregrounds.

\section{Conclusions}
Our analysis of the \cbi\ observations has yielded parameters
consistent with the standard $\Omega_{tot}=1$, $\Lambda$CDM
model. These results, based on measurements extending to much higher
$\ell$ than previous experiments, provide a unique confirmation of the
model. The dominant feature in the data is the decline in the power
with increasing $\ell$, a necessary consequence of the paradigm which
has now been checked. In summary under weak prior assumptions the
combination of \cbi\ and \DMR\ data gives $\Omega_{tot} =
1.01_{-0.06}^{+0.09}$, and $n_s = 1.02_{-0.07}^{+0.11}$, consistent
with inflationary models; $\Omega_{\rm cdm}h^2=0.12\pm0.03$, and
$\Omega_{\Lambda} = 0.64_{-0.14}^{+0.11}$. With more restrictive
priors, flat+weak-$h$+LSS, are used, we find $\Omega_{\rm
cdm}h^2 = 0.13_{-0.01}^{+0.02}$, consistent with large scale structure
studies; $\Omega_b h^2 = 0.025_{-0.008}^{+0.010}$, consistent with Big
Bang Nucleosynthesis; $\Omega_m = 0.37\pm 0.11$, and
$\Omega_b = 0.060\pm 0.020$, indicating a low matter density universe;
$h = 0.65_{-0.12}^{+0.12}$, consistent with the recent determinations
of the Hubble Constant based on the recently revised Cepheid
period-luminosity law; and $t_0=14.0_{-1.2}^{+1.2}$ Gyr, consistent
with cosmological age estimates based on the oldest stars in globular
clusters. The combination of CMB measurements and LSS priors also
enables us to constrain the normalization $\sigma_8$. We find that for
flat+weak-$h$+LSS priors we obtain
$\sigma_8=0.89_{-0.10}^{+0.14}$. Thus it appears that the
normalization required to explain the excess with the SZE is in the
upper range of the independent result based on the primary CMB signal
and LSS data.

The 2001 observing season data is now being analyzed. Although the
data will double the overall integration time it is not expected to
increase the confidence of the high-$\ell$ measurements as the
observations were aimed at doubling the mosaic area and not at further
integration of the deep fields. Follow-up surveys of the deep fields
in the optical range and correlation with existing X-ray catalogs
may establish whether the measurement is indeed a serendipitous
detection of the SZE and will be part of future work. However, the
observations have highlighted the potential for SZE measurements to
constrain $\sigma_8$ via the highly sensitive dependence of the
angular power spectrum to the amplitude of the fluctuations ${\cal
C}^{SZ}\sim\sigma_8^7$, although precise calibration of the theories
from either numerical or analytical methods are required to make such
conclusions feasible\cite{Bond02}. The \cbi\ is currently being
upgraded with polarization sensitive antennas for the 2002/2003
observing season.

This work was supported by the National Science
Foundation under grants AST 94-13935, AST 98-02989, and AST
00-98734. Research in Canada is supported by NSERC and the Canadian
Institute for Advanced Research. The computational facilities at
Toronto are funded by the Canadian Fund for Innovation.  We are
grateful to CONICYT for granting permission to operate the CBI at the
Chajnantor Scientific Preserve in Chile.

\end{document}